\documentclass[]{aiaa-tc}

\usepackage{mathtools}

\usepackage{epstopdf}
\usepackage{bm}  
 \title{Wind Turbine Large-Eddy Simulations on 
 Very Coarse Grid Resolutions 
 using an Actuator Line Model}

 \author{
  Luis A. Mart\'inez Tossas 
  \thanks{Ph.D.~Candidate, Mechanical Engineering, Johns Hopkins University, AIAA Member.}
  , \ Richard J.A.M. Stevens
    \thanks{Post-Doctoral Fellow, Mechanical Engineering, Johns Hopkins University and 
    Department of Physics, Mesa+ Institute, and J. M. Burgers Centre for Fluid Dynamics, University of Twente.}
  \ and \ Charles Meneveau 
  \thanks{L.M. Sardella Professor of Mechanical Engineering, Johns Hopkins University, AIAA Member.}
  \\
  {\normalsize\itshape
   Johns Hopkins University, 
   Baltimore, MD, 21210, USA}\\
 }

 \AIAApapernumber{YEAR-NUMBER}
 \AIAAconference{Conference Name, Date, and Location}
 \AIAAcopyright{\AIAAcopyrightD{YEAR}}

\begin{document}

\maketitle

\begin{abstract}
In this work the accuracy of the Actuator Line Model (ALM) in Large Eddy Simulations of wind turbine flow is studied
under the specific conditions of very coarse  spatial resolutions. 
For finely-resolved conditions, it is known that ALM provides  better accuracy compared to  the 
standard  Actuator Disk Model (ADM) without rotation.
However, we show here that on very coarse resolutions, flow induction occurring at rotor 
scales can affect the predicted inflow angle and can
adversely affect the ALM predictions.
We first provide an illustration of coarse LES to reproduce wind tunnel measurements. 
The resulting flow predictions are good, but the challenges 
in predicting power outputs from 
the detailed ALM motivate more detailed analysis on a 
case with uniform inflow.
We present a theoretical framework to 
compare the filtered quantities that enter the Large-Eddy Simulation equations as body forces with a scaling relation
between the filtered and unfiltered quantities. 
The study aims to apply the theoretical derivation to the simulation framework and improve
the current results for an ALM, especially in the near wake where the largest differences are observed.

\end{abstract}



\section{Introduction}

Wind farm simulations require intensive 
computational resources and advanced 
turbine models in order to capture rich
amounts of information such as loads along the blades,
aerodynamic forces, and power output accurately
\cite{churchfield_large-eddy_2012, martinez-tossas_large_2014}.
The Actuator Line Model (ALM) is known to
depend on a high grid resolution 
(on the order of 50 grid points per rotor) 
when using Large-Eddy Simulations (LES)
to provide grid-converged results
\cite{sorensen_numerical_2002,martinez-tossas_large_2014}.
However, in order to study  large wind farms and the associated large-scale properties of coupling with
the atmospheric boundary layer, wind farms consisting of many turbines must be considered \cite{Calafetal10,Stevensetal14a,Stevensetal14b}.
When many turbines must be included in the computational domain, fine spatial resolution on each rotor is often not affordable, especially if one wishes to repeat the LES varying flow conditions, rotor designs, etc.  It is thus of interest to study and better understand the performance of the ALM approach when using very coarse discretization (on the order of 10 LES grid-points per rotor diameter).  

It has been seen that power extracted from an ALM in LES is very dependent
on the grid resolution and the width
of the kernel used to smear the forces \cite{martinez-tossas_large_2014}.
The wake structures are also affected by means of the spatial filter
in the velocity field which is not able to resolve the structures 
smaller than the grid size.
The subgrid-scale closure of the LES models the smaller scales of the flow,
but because of the spatially filtered velocity
may differ from the real unsmoothed velocity,
differences of predicted forces and power at the
rotor blade can exist.
The axial induction can be affected by the smearing
of the velocity field.

Here, we study the implementation of an  ALM in
coarse simulations (14 LES grid-points per rotor).
First, we examine the performance of the ALM model for a 
single wind turbine with turbulent inflow conditions for which
detailed wind tunnel experimental data are available. 
Second, a formulation is presented to address differences in
power caused by the width of the kernel being used.
A single turbine in uniform inflow is studied to document 
the effect of the
smearing kernel on blade loads and power production.

\section{Numerics}

\subsection{Solver}
The current study uses LESGO, \cite{Calafetal10,Stevensetal14b}
the Johns Hopkins Turbulence Research Group's  Large-Eddy Simulations framework 
including a recent implementation of the ALM. \cite{martinez_wake_2015}
LESGO solves the filtered Navier-Stokes Equations
for very large Reynolds number
\begin{equation}
    \frac{\partial \widetilde{{\textbf{u}}}} {\partial t} 
    +(\widetilde{{\textbf{u}}} \cdot \nabla) \widetilde{{\textbf{u}}}
    =-\frac{1}{\rho}\nabla \widetilde{p} + 
    -\nabla \cdot {\bm {\tau}}
    + \frac{\widetilde{\textbf{f}}}{\rho},
    \label{eq:NS}
\end{equation}
where $\widetilde{~}$ denotes spatial filtering,
$\textbf{u}$ is the velocity,
${p}$ is pressure,
$\rho$ is density, 
$\bm \tau$ is the subgrid scale stress tensor,
and 
$\textbf{f}$ is the body force.
The code uses pseudo-spectral representation  in the two horizontal directions 
with the third (vertical) direction using 
second-order centered finite differencing. The approach is kinetic-energy conserving. 
Time advancement is done using a second order 
Adams-Bashforth scheme. 
Subgrid-scale modeling is based on an
eddy viscosity type 
scale-dependent Lagrangian dynamic model \cite{bouzeid_2005}.
The model represents the deviatoric part of the SGS stress tensor,
$\tau^d_{ij}$, as
\begin{equation}
  \tau^d_{ij} = 
  -2 \nu_{SGS} \widetilde{S}_{ij} = 
  -2 (C_S \Delta)^2 |\widetilde{\bm S}| \widetilde{S}_{ij},
\end{equation}
where $C_S$ is the dynamically computed 
Smagorinsky coefficient,
$\Delta$ is the grid resolution, 
and $\widetilde{S}_{ij}$ is the 
symmetric part of the resolved velocity gradient tensor \cite{pope_2000}.

\subsection{Actuator Line Model (ALM) and Turbine Design}
The ALM implementation in LESGO \cite{martinez_wake_2015}
incorporates a wind turbine into the Navier-Stokes Equation,
see Eq. \ref{eq:NS},
as a body force term \cite{sorensen_numerical_2002}.
This body force is calculated dynamically depending
on the local velocity at the location of each actuator point.
The actuator points are used to calculate the blade forces with 
a high accuracy before they are projected on the 
computational grid used to calculate the flow field.
Lift and drag are calculated based on
tabulated airfoil data and then smeared onto the flow field 
as body forces by means of a Gaussian kernel
\begin{equation}
  \eta_\epsilon = \frac{1}{\epsilon^3 \pi^{3/2} } 
                  e^{-\left( r^2 / \epsilon^2 \right)},
\end{equation}
where $r$ is the distance from the actuator point,
and $\epsilon$ establishes the width of the kernel.
The two turbines used for this study are the
turbine used in the experiments from the group
at \'Ecole Polytechnique F\'ed\'erale de Lausanne (EPFL)
which has a 0.15 m diameter, \cite{wu_simulation_2012}
and the NREL 5MW Reference,
which has a 126 m rotor diameter \cite{jonkman}.

\begin{figure}[htb!]
\centering
\includegraphics[width=0.4\linewidth]{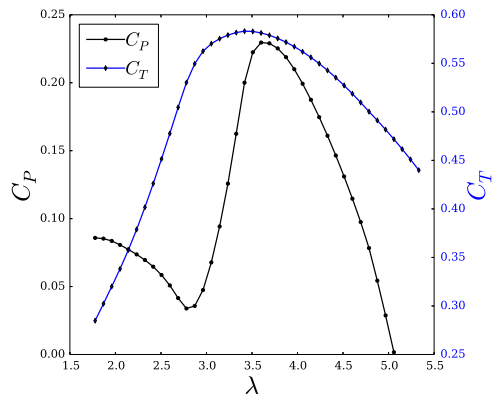}
\caption{Power and thrust coefficient as a function of tip speed ratio
for a GWS/EP-6030
computed using Blade Element Momentum Theory.}
\label{fig:cpct}
\end{figure}

The turbine used by the group at 
EPFL
is a GWS/EP-6030. \cite{wu_simulation_2012, wu2010}.
The chord and pitch angle along the blades are 
found in their previous work \cite{wu2010}.
The lift and drag coefficient tables were obtained by
using a cambered airfoil,
i.e. we used airfoil (9) from the work by Sunada et al. \cite{sunada}
In order to match the wake profiles close to the center of the
wake, a nacelle has to be included in the model.
Power and thrust coefficient curves have been
obtained using Blade Element Momentum Theory and
are shown in Figure \ref{fig:cpct}.
It can be observed that this turbine provides a poor
performance in terms of power, i.e. low $C_P$ value,
compared to a large scale turbine which
operates on a range closer to the Betz limit.
The thrust coefficient, however, is closer to typical values
observed for operational large scale turbines.

\section{Results}

\subsection{Comparisons with wind tunnel data with turbulent inflow}
We now present simulation of a wind turbine
with turbulent inflow. 
We use the experiment from the EPFL group \cite{wu_simulation_2012}, 
here focusing on the single-turbine case.
The inflow condition is matched by using a 
concurrent-precursor simulation
matching the given roughness height and 
friction velocity of the experiment 
\cite{wu_simulation_2012}.
The tip speed ratio of the turbine is set to $\lambda=4$ as in the experiment.
As a first step we compare the current implementation of the 
actuator line model (ALM) with and without the effects of the nacelle.
Here the LES uses a coarse discretization, 14 points across the rotor.
The ALM has been tested with and without a nacelle and tower modeled as drag objects 
with $C_D$ coefficients of 1.0 and 0.5 respectively.

\begin{figure}[htb!]
\centering
\includegraphics[width=0.99\linewidth]{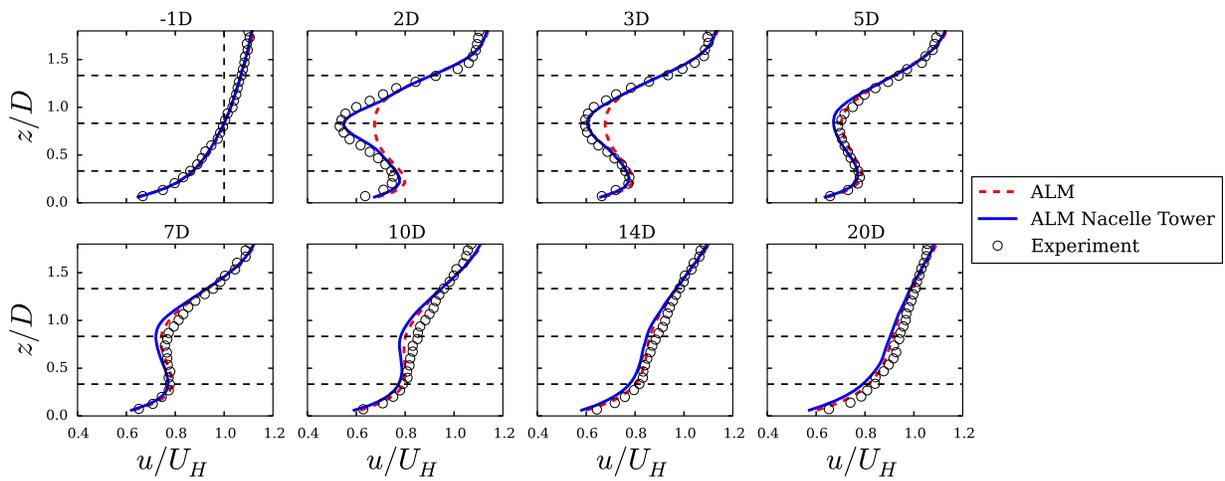}
\caption{Wake profiles for a single turbine in the EPFL wind 
tunnel experiment. $U_H$ is the mean velocity at hub-height
at the inflow location (-1D).}
\label{fig:multiplotU}
\end{figure}

\begin{figure}[htb!]
\centering
\includegraphics[width=0.99\linewidth]{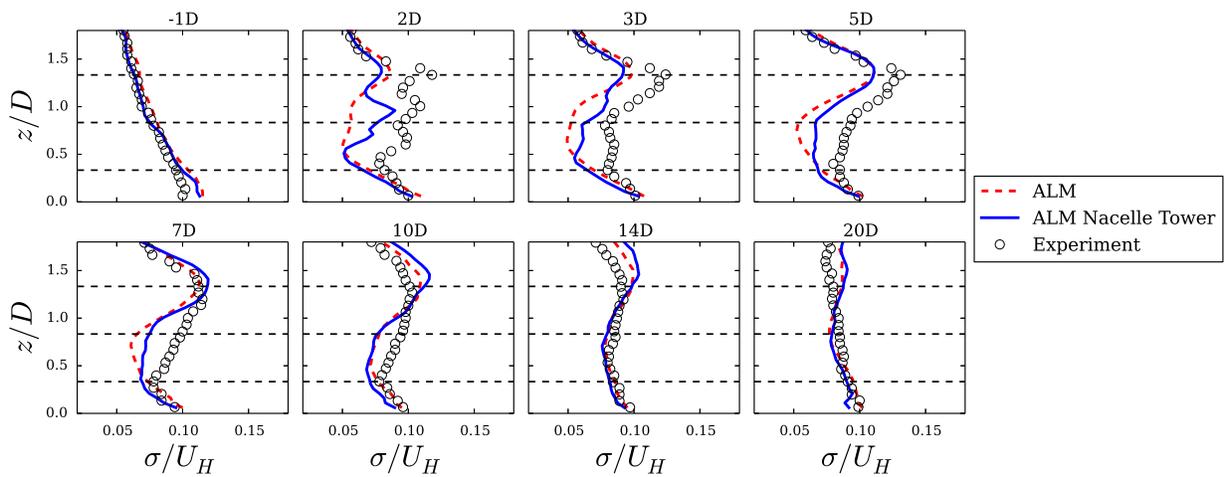}
\caption{Turbulence intensity profiles for a single turbine in the EPFL wind 
tunnel experiment.}
\label{fig:multiplotupup}
\end{figure}

Figures \ref{fig:multiplotU} and \ref{fig:multiplotupup} show 
streamwise velocity and turbulence intensity
profiles across the wake for a single turbine in a turbulent boundary layer.
The non-dimensional scale used is the hub-height velocity $U_H$.
It is to be noted that in order to obtain good agreement between the
experiments and simulations a nacelle model needs to be included.
The current implementations provides excellent agreement
for the velocity profiles of the wake.
The turbulence intensity computed in the LES under-predicts
the experimental measurements.
This is expected from LES, since some of the scales are filtered,
thus reducing the fluctuations.
The overall trends agree quite well, and it can be observed
that at 2-3D downstream the wake of the nacelle and tower have a
big influence
in both the mean velocity and turbulence intensity.

The wake profiles obtained agree very well with the experiments.
However, quantities at the turbine, such as power and thrust,
are very dependent on the grid resolution
and the value of the kernel width $\epsilon$. \cite{martinez-tossas_large_2014}
The power coefficient obtained from the simulation
for this turbine is $C_P=0.3$.
This power coefficient is significantly higher 
than what is predicted
by the turbine design using BEM (Figure \ref{fig:cpct}).
While the overall thrust may be realistic
(since the wake profiles agree well),
the differences between the predicted $C_P$
and the expected $C_P$ from BEM points to
possible problems due to the smearing of velocity inherent 
in ALM with very coarse LES resolutions.
In order to better understand the discrepancy, first we perform 
a simulation under uniform inflow, 
and then present some initial thoughts of how
the model inaccuracies may be remedied.

\subsection{Simulations of NREL turbine with uniform inflow}
\label{sec:5MW}
\begin{figure}[htb!]
\centering
\includegraphics[width=0.4\linewidth]{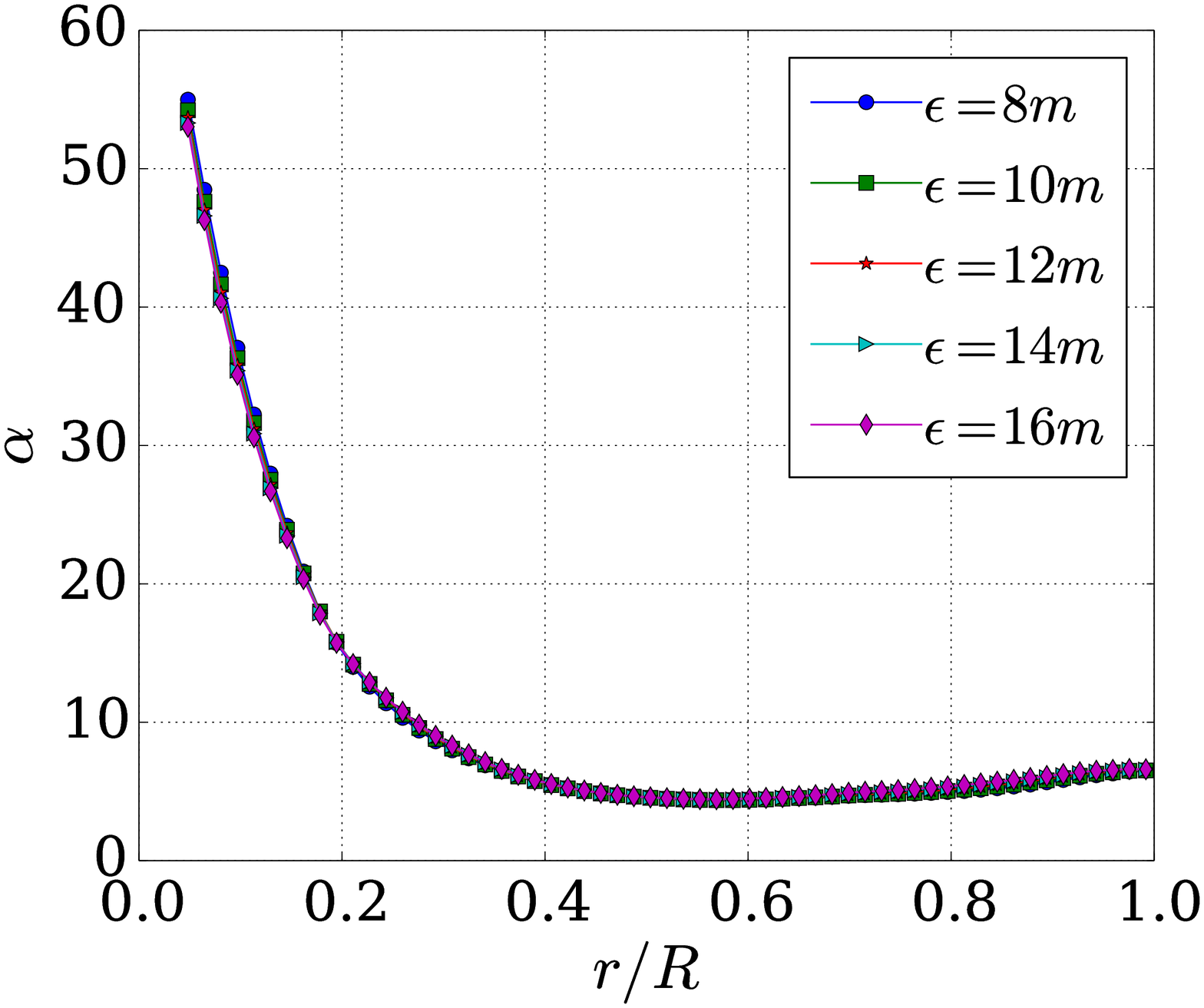}
\includegraphics[width=0.4\linewidth]{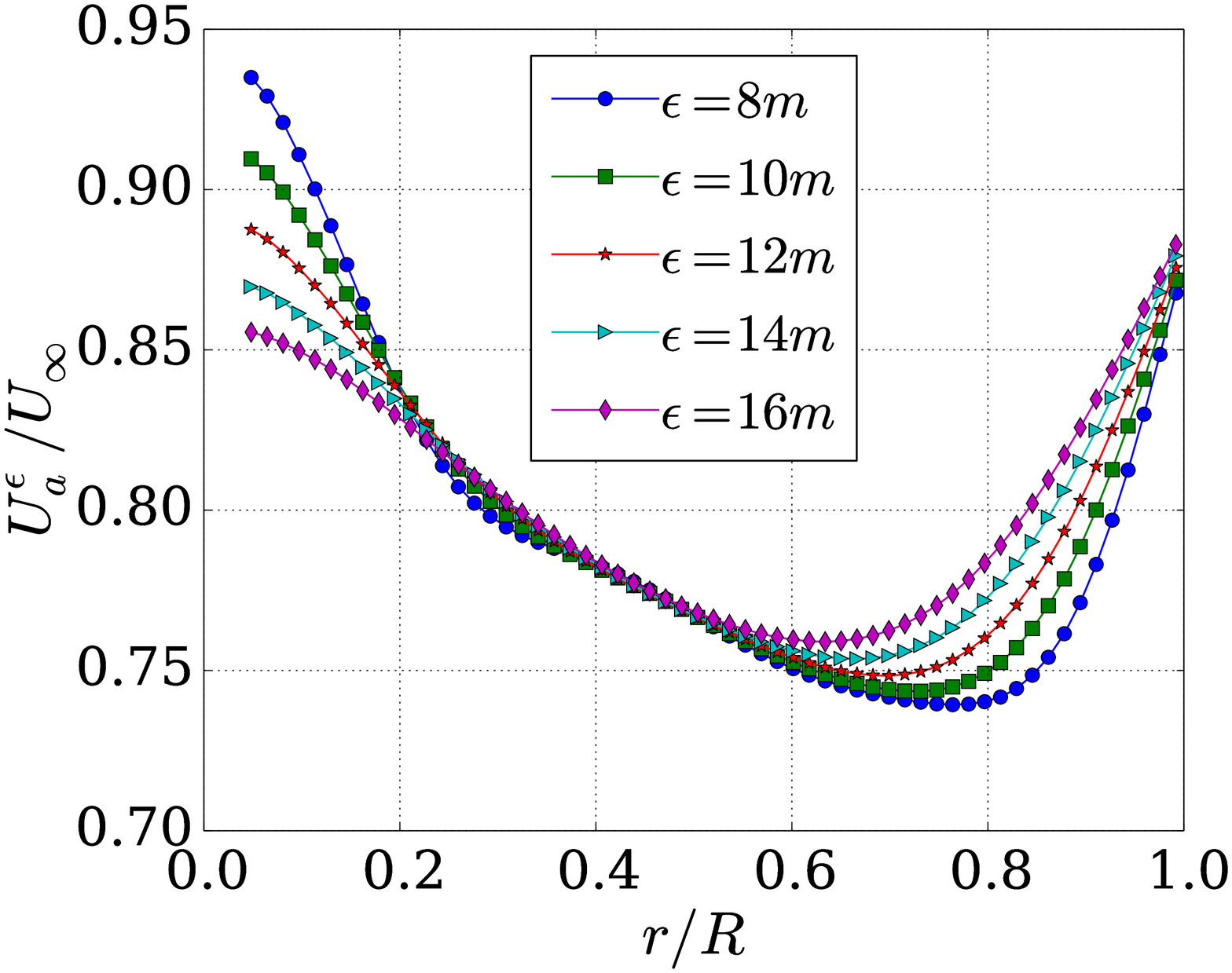}

\includegraphics[width=0.4\linewidth]{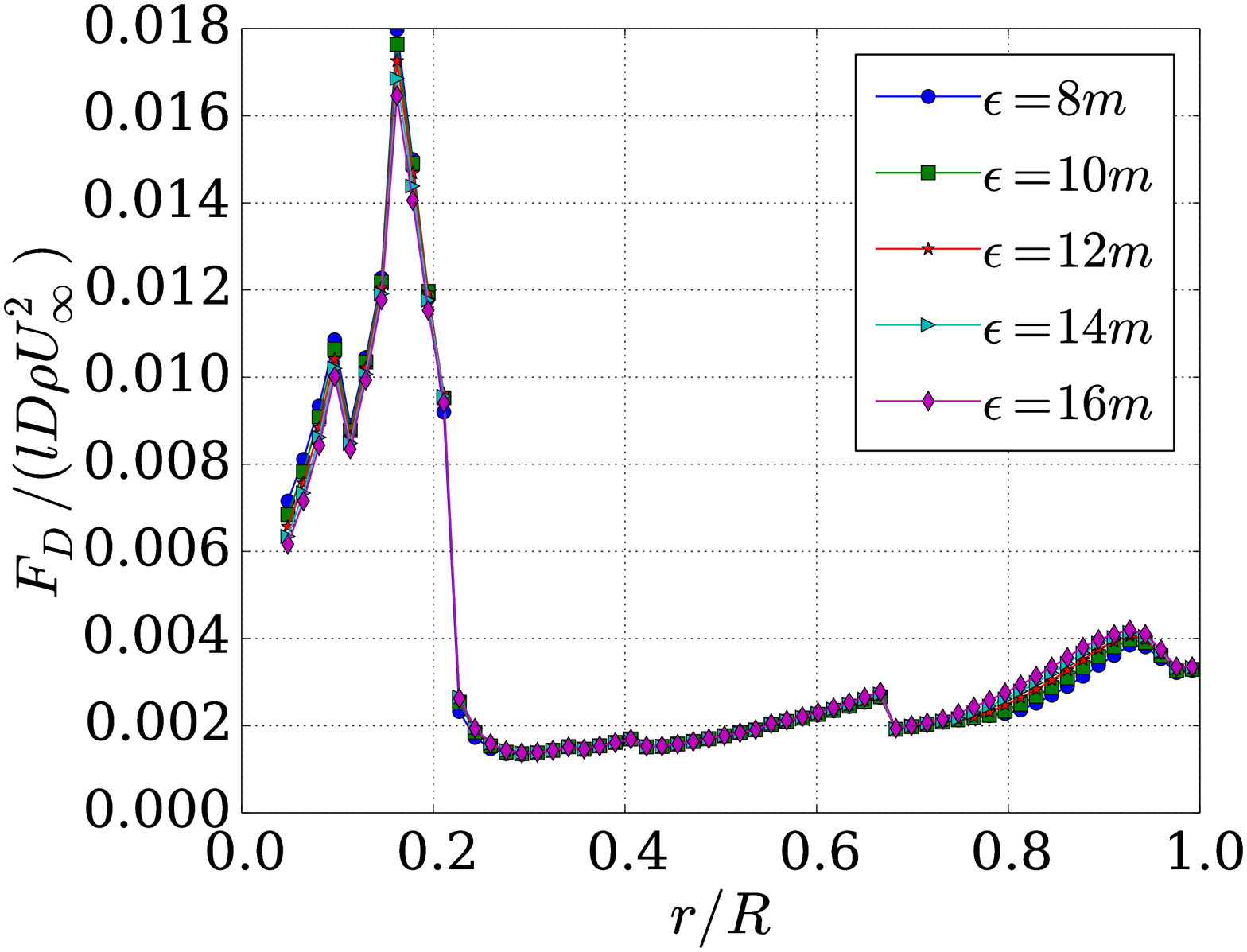}
\includegraphics[width=0.4\linewidth]{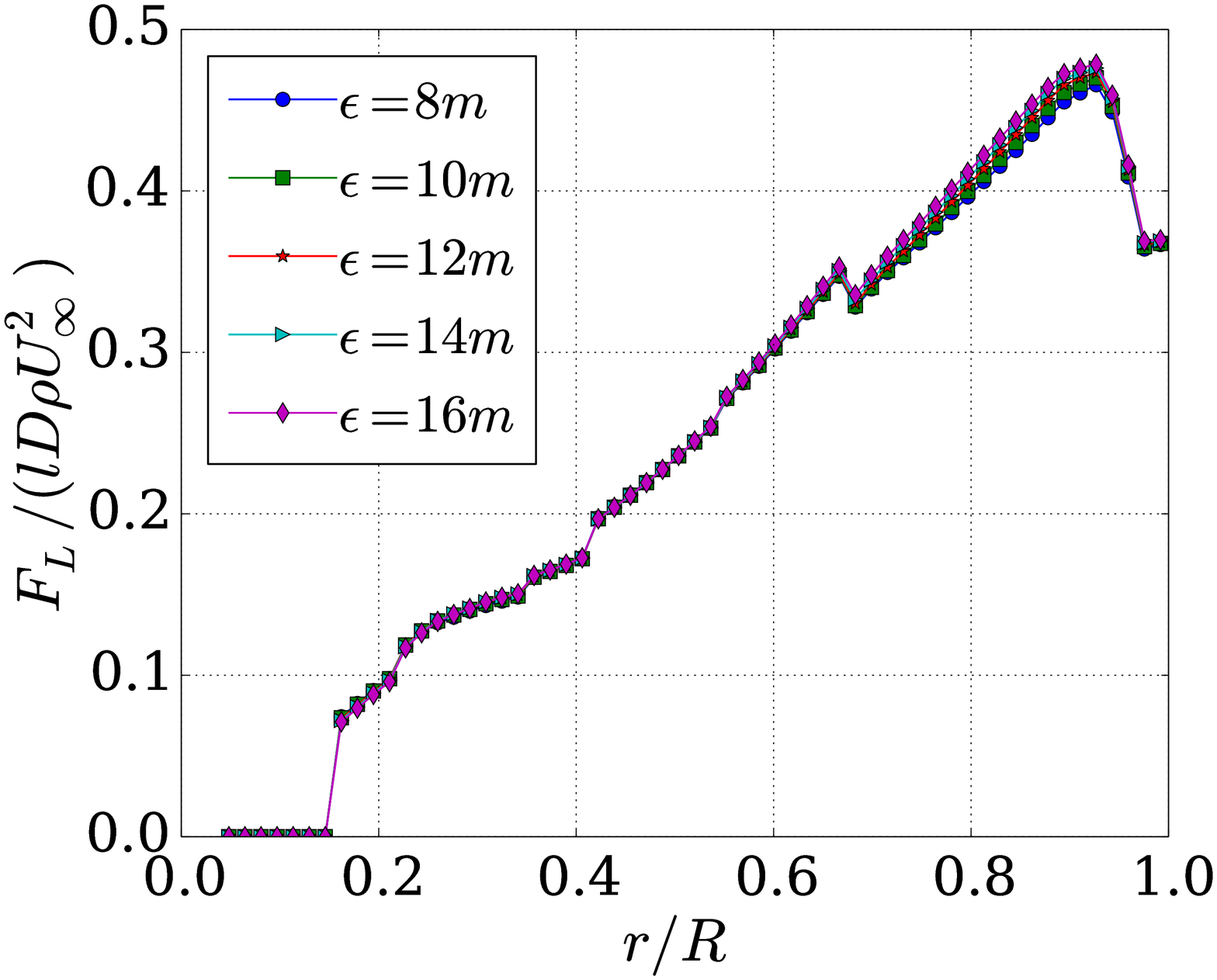}
\caption{(a) Angle of attack, (b) axial velocity, (c) normalized drag force, and (d) normalized lift force as function of the blade length. The data points coincide with the actuator points used in the simulation.}
\label{fig:bem}
\end{figure}

We consider the simulation of the NREL wind turbine with laminar uniform inflow. 
We use LES with a grid resolution of $\Delta = $ 4 m in each direction, 
an inflow velocity of 
8 (m/s), a prescribed tip-speed ratio of $\lambda = 7.5$, 
and vary the Gaussian kernel width $\epsilon$. 
It is known that wind turbine loads along the blades depend significantly 
on the filtering size of the Gaussian kernel used to smear the forces
of the turbine. Figure \ref{fig:bem} shows  angle of attack, axial velocity, lift and drag
along the blade for an NREL 5MW Reference turbine modeled
as an actuator line with different kernel widths ($\epsilon$).
From this figure we can see that the largest difference 
is in axial velocity. This difference will cause differences in the 
lift and drag curves, that are more pronounced as
we move further towards the tip. These differences change the total
power that the turbine is extracting from the flow.

\subsection{Effects of filtering}

\label{sec:derivation}

Next, we present some analytical considerations focusing on the effects of the spatial filtering 
inherent in LES and its impacts on the ALM computations of forces. 
We take the lift component of the body force term per unit width 
in a 2D frame
and write it (simplifying) as:
\begin{equation}
\widetilde{f} = \frac{1}{2} c \rho \widetilde{C_L U^2}
\end{equation}
where $c$ is the chord, $\rho$ is the density,
$C_L$ is the lift coefficient, and $U$ is the
magnitude of the local velocity
on the reference frame of the moving airfoil section.
The $C_L$ coefficient is a function of $U$ through its dependence on angle of attack.
For small angle of attack, $sin(\alpha) \approx \alpha$ and thus the angle of attack 
can be expressed as $\alpha \approx U_a / U$, where
$U_a$ is the axial velocity.
The lift coefficient for small angles of attack is given by
$C_L = B (\alpha + \alpha_0)$ where $B$ is a constant and
$\alpha_0$ is an offset.
We will assume $\alpha_0 << \alpha$ so that
\begin{equation}
C_L = B \alpha = B \sin^{-1}\left( \frac{U_a}{U} \right)
\approx B \frac{U_a}{U}.
\end{equation}
The filtered body force can be defined as
\begin{equation}
\widetilde{f} = \frac{1}{2} c \rho 
B\widetilde{ \frac{U^2U_a}{U}} =  \frac{1}{2} c \rho B  \widetilde{U U_a }
\end{equation}
Now we can also write a similar equation for the unfiltered force:
\begin{equation}
 f = \frac{1}{2} c \rho B  U U_a 
\end{equation}
and from here:
\begin{equation}
f = \widetilde{f} 
\left(
\frac{U U_a}{\widetilde{U U_a }}
\right)
\end{equation}
If we use the argument that the velocity magnitude $U$ is dominated by  
the rotational component of the velocity vector (which is valid for high TSR's)
then we can take the velocity magnitude out of the filtering operation and we obtain
\begin{equation}
f \approx \widetilde{f} 
\left(
\frac{U_a}{\widetilde{U_a }}
\right)
\label{eq:ratio}
\end{equation}
The relationship in Eq.  \ref{eq:ratio} enables us to interpret the results shown in 
Fig. \ref{fig:bem} (top right) in the sense  that 
the ratio of unfiltered force equals the ratio
of unfiltered to filtered {\it axial} velocity.
We wish to determine the unfiltered force 
but only know the filtered force.
If we can model ratio of unfiltered and filtered axial velocity,
Eq \ref{eq:ratio} would enable us 
to determine the correct (unfiltered) force.

While the derivation of Eq. \ref{eq:ratio} was based on spatial filtering, we now interpret variables filtered at a scale $\epsilon$ as the variables one obtains by running LES and ALM with a kernel width $\epsilon$. While this assumption cannot be firmly established on theoretical grounds, in practice it is a useful assumption and is commonly made in LES.  For example, if we denote the axial velocity and force obtained in a simulation with a kernel $\epsilon$ as $U_a^\epsilon$ and $f^\epsilon$, and associate the scale $\epsilon$ with the filtering scale, then  $U_a^\epsilon \approx \widetilde{U_a}$ and $f^\epsilon \approx \widetilde{f}$. As a next step, we filter Eq. \ref{eq:ratio} 
at another filter scale $\epsilon_{min} < \epsilon$ and assume that the $\epsilon$-variables are not affected by this smaller-scale filtering. Then one can rewrite Eq. \ref{eq:ratio} according to 
\begin{equation}
U_a^{\epsilon_{\rm min}*} = f^{\epsilon_{\rm min}} \times \frac{ U_a^{\epsilon} }{f^\epsilon}. 
\label{eq:ua2}
\end{equation}
We can use simulations performed at various $\epsilon$'s to test this relationship, as a first step in establishing possible correction strategies for the ALM approach. 

\begin{figure}[htb!]
\centering
\includegraphics[width=0.4\linewidth]{figures/epsilon/Vaxial.eps}
\includegraphics[width=0.4\linewidth]{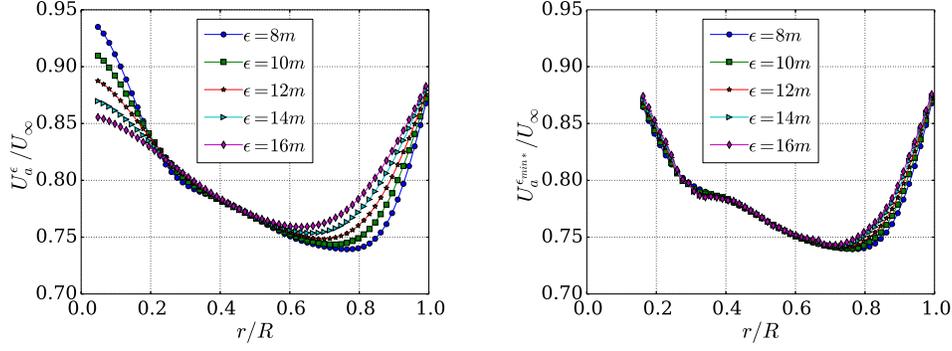}
\caption{Axial velocity for different $\epsilon$ values (left) and
re-scaled axial velocity (right) 
to $\epsilon=$ 8m case as a 
function of blade length,
where $U_a^{\epsilon_{min}*} = f^{\epsilon_{min}} \times \frac{ U_a^{\epsilon} }{f^\epsilon}$ and 
$\epsilon_{min} = 8$ m.
The data points coincide with the actuator points used in the simulation.}
\label{fig:ua}
\end{figure}

Returning to our simulations of the NREL case, we consider $\epsilon_{\rm min} = 8$m, and attempt to use the results from the other 4 runs with $\epsilon>\epsilon_{\rm min}$ to rescale the force $f^{\epsilon_{\rm min}}$. If Eq. \ref{eq:ua2} holds, 
the distribution of $U_a^{\epsilon_{\rm min}}$ 
should coincide with  the RHS of Eq. \ref{eq:ua2},
i.e. with $U_a^{\epsilon_{\rm min}*}$, regardless of $\epsilon>\epsilon_{\rm min}$ chosen.
The first region of the blade is made of cylindrical
sections which produce no lift, thus the 
approach presented is not used in this section.
As seen in Figure \ref{fig:ua}
the curves collapse through most of the blade except
close to the tip.
The reason for this is that near the tip, 
the 3D effects
of the kernel become more noticeable.

\section{Conclusion}
The ALM has been validated on a relatively coarse resolution
with experimental measurements from the group at
EPFL \cite{wu2010,wu_simulation_2012}.
The LES provides good agreement with the experiments,
but a nacelle and tower model are required to obtain the proper
wake deficit.
When analyzing power output computed by the ALM,
the values are higher than those predicted 
by Blade Element Momentum Theory.
A theoretical analysis is presented that attempts to
relate filtered forces
to filtered velocities,
highlighting the impact of filtering of the 
axial velocity on the force distribution.
The derived relationship is tested based on 
simulations done at various
$\epsilon$ values and is seen to hold rather well,
except when approaching the tip of the blade.
The results imply that one may be
able to predict the fine-grained
force corresponding to an $\epsilon$ smaller than
that of the simulation,
if one were able to deduce the corresponding
axial velocity 
from the data at coarser $\epsilon$.
Such a model-based defiltering 
operation is the focus of ongoing investigations.

\section*{Acknowledgments}

LAMT was supported by NSF (IGERT), while international travel and CM were supported by NSF (IIA 124382, the WINDINSPIRE project). 
RJAMS is supported by the research program Fellowships for Young Energy Scientists (YES!) of the Foundation for Fundamental Research on Matter (FOM) supported by the Netherlands Organization for Scientific Research (NWO).

\end{document}